\documentclass[final,3p,twocolumn,times]{elsarticle}
\usepackage{amsmath}
\usepackage{amsfonts}
\usepackage{amssymb}
\usepackage{bm}
\usepackage{booktabs}
\usepackage{xcolor}
\usepackage{subfig}
\usepackage{graphicx}
\usepackage{url}
\usepackage{adjustbox}
\usepackage{pdflscape}
\usepackage{afterpage}
\usepackage{booktabs}
\usepackage{enumitem}
\usepackage{makecell}
\usepackage{comment}
\usepackage{balance}
\usepackage{color}
\usepackage{soul}
\usepackage{mwe,tikz}
\usepackage[percent]{overpic}

\DeclareMathOperator{\ingeo}{\prec}

\journal{}

\begin{document}

\begin{frontmatter}
\title{A worldwide study on the geographic locality of Internet routes}

\author[dip]{Massimo Candela}
\ead{massimo.candela@ing.unipi.it}
\author[iit]{Valerio~Luconi}
\ead{valerio.luconi@iit.cnr.it}
\author[dip]{Alessio~Vecchio\corref{cor1}}
\ead{alessio.vecchio@unipi.it}

\cortext[cor1]{Corresponding author}
\address[dip]{Dip. di Ing. dell'Informazione, Universit\`a di Pisa, Largo L. Lazzarino 1, 56122 Pisa, Italy}
\address[iit]{Istituto di Informatica e Telematica, Consiglio Nazionale delle Ricerche, Via G. Moruzzi, 1, 56124 Pisa, Italy}

\date{March 2021}

\begin{abstract}
The topology of the Internet and its geographic properties received significant attention during the last years, not only because they have a deep impact  on the performance experienced by users, but also because of legal, political, and economic reasons. In this paper, the global Internet is studied in terms of path locality, where a path is defined as local if it does not cross the borders of the region where the source and destination hosts are located. 
The phenomenon is studied from the points of view of two metrics, one based on the size of the address space of the autonomous systems where the endpoints are located and the other one on the amount of served population. Results show that the regions of the world are characterized by significant differences in terms of path locality. The main elements contributing to the path locality, and non-locality, of the regions and countries, are identified and discussed. Finally, we present the most significant dependency relationships between countries caused by non-local paths.
\end{abstract}

\end{frontmatter}

\section{Introduction}

The geographic properties of the Internet have been the subject of numerous studies during the last years not only for technical reasons, but also because of economic, societal, and geopolitical ones~\cite{bachmann2017improving,malecki2009wired,6614151,CANDELA2020107495}. 
The connection between the economic world and the geographic properties of the Internet goes beyond routing costs. 
The ongoing clustering of production facilities, as well as the increasing urban agglomeration, play a role in the policies behind the design of Internet infrastructure: anthropic activities, and their associated economic value, are the forces driving the Internet evolution~\cite{doi:10.1111/j.1944-8287.2002.tb00193.x}. At a large scale, the uneven economic development of countries is reflected by the heterogeneous technical advancements of the global Internet infrastructure~\cite{doi:10.1080/09654310701748009}. All these factors play a role in defining the geographic properties of Internet paths, including the set of traversed countries. The position of Internet eXchange Points (IXPs), server farms, and Points of Presence (PoPs) is influenced by the anthropic and economic geography of a region.   

In some circumstances, geopolitical factors may be more important than the economic ones in defining the geographic properties of the Internet. Network infrastructure is increasingly considered by governments as a possible target of hostile countries, pushing towards the definition of borders also in cyberspace. Authoritarian governments try to limit the freedom of expression made possible by the Internet by exerting control within their sphere of influence~\cite{deibert2009geopolitics}. Even in the absence of friction, some national regulations may establish a connection between the Internet and geography. A notable case is represented by the regulations about data protection and privacy for all the individual citizens of the EU, which also regulates the transfer of personal data outside the Union~\cite{gdpr}. 

\begin{figure} 
\centering
\includegraphics[width=0.9\columnwidth]{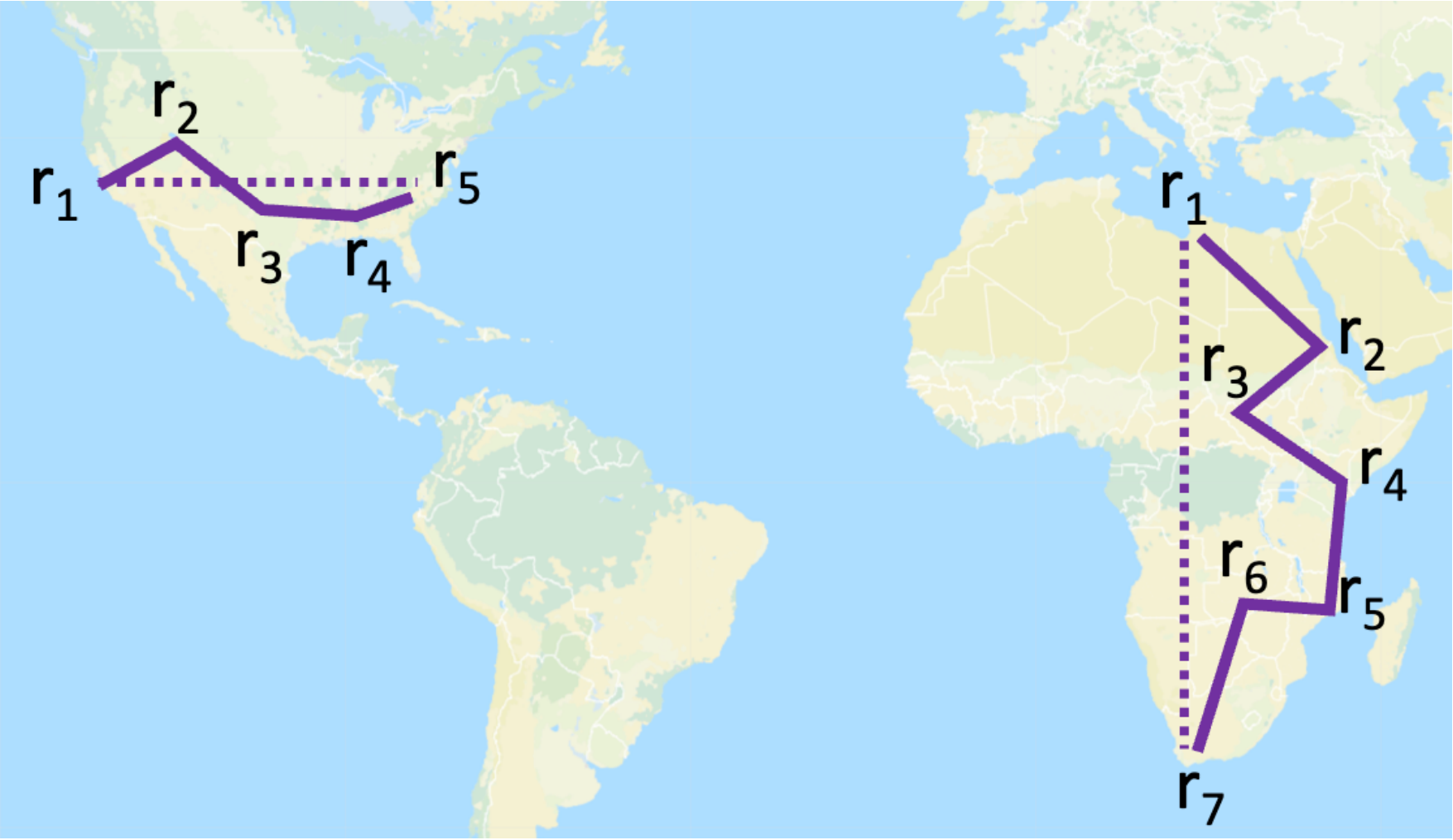}
	\caption{The path in the USA is characterized by a relatively reduced amount of circuitousness, the path in Africa is more circuitous; the dashed line represents the ideal -- and shortest -- path between source and destination. 
	\label{fig:circuitousness}}
\end{figure}

From a technological perspective, observing which is the physical path that is taken by packets allows a better understanding of the geographic efficiency of routes, where, in this specific case, efficient means characterized by a reduced amount of circuitousness~\cite{1217279, MATRAY20122237}. Circuitousness can be defined as the amount of deviation from the shortest path on the surface of the earth. More formally, given a path $p=\{r_1, r_2, ..., r_k\}$ at the IP level, its circuitousness $c(p)$ can be defined as 

\[ c(p) = \frac{\sum_{i=1}^{k-1} d(r_i, r_{i+1})}{d(r_1, r_k)} \]

\noindent where $d(r_i, r_j)$ is the geographic distance between $r_i$ and $r_j$, and $k$ is the number of elements in the path. In the real world, $c(p)$ is typically greater than 1: source and destination are generally not directly connected and the presence of intermediate routers makes the path longer than the ideal -- and shortest -- one. Figure~\ref{fig:circuitousness} shows two paths: the one in the USA is slightly circuitous, whereas the one in Africa is much more circuitous.

Circuitousness may significantly contribute to the end-to-end delay. There is an always growing number of applications that are extremely sensitive to latency, for instance streamed video games, VoIP, or remote-controlled systems~\cite{SINGH2014365, singla2014internet}. For any application, the propagation delay may represent a significant fraction of the overall communication latency, especially when the route spans over a geographically extended region. Routing is generally characterized by reduced circuitousness within the boundaries of a single Autonomous System (AS). In particular, Nur et al. found that the ingress-to-egress subpaths have lower circuitousness than the end-to-end paths~\cite{Nur:2018:GRI:3276978.3239162}. This indicates that the infrastructure and the routing schemes adopted by the single ASes are generally efficient. However, this does not apply necessarily when considering routing on a global scale: in inter-AS routing, ASes may prefer locally optimized routes instead of globally optimized ones, or they may just lack awareness regarding this aspect.

In this paper, we study the locality of Internet paths. A path is local if it does not cross the boundaries of the area where the source and destination hosts are located. From the background and scenarios discussed above, it should be clear that a high level of path locality for an area can be: i) evidence of reduced technological dependence from external parties, ii) a possible element for evaluating, on a large scale, compliance to regulations, iii) an indicator of topologically efficient communication networks, as non-local paths are implicitly circuitous. 
We evaluated the path locality of both continent-scale areas and country-scale ones.
The path locality of all the considered areas has been determined using a large dataset of Internet measurements. In particular, measurements were collected by means of RIPE Atlas, the most extensive public measurement platform actually available. The presented study focuses on the technical side of path locality: we define two path locality metrics, we identify the main issues in collecting the necessary data, we present and discuss the results obtained on a global scale and for the different regions of the world at the infrastructure level. Some economic and geopolitical considerations are also provided, albeit they are not the main analysis criteria. Results show that the regions of the world are characterized by different levels of path locality: the paths of Europe, North America, and Oceania are almost always local; on the contrary, Africa, Asia, the Middle East, and South America show some dependence on external communication infrastructure.

\section{Related work and contribution}
\label{sec:related}

In the following, we summarize the most significant work related to the geographic extension of the Internet infrastructure, IP geolocation, and path locality.

\subsection{Internet and geography}

Methods for the discovery of the Internet topology and its global-level performance received significant attention during the last decade~\cite{7076582,choffnes2010crowdsourcing,gregori2013sensing,6710071,GREGORI201662}.
Besides its topological structure, also the geographic dimension of the Internet has been the subject of investigation~\cite{10.1145/964723.383073}. The position of routers can be used, in fact, to characterize the relationship between population and network infrastructure density, the distribution of link lengths, and the extent of ASes~\cite{1217279}.

The geographic properties of Internet routing were analyzed in several studies, which highlighted the presence of  circuitous paths \cite{subramanian2002geographic, 5934915}. 

Another work, about the relationship between round-trip time (RTT) and geography, highlighted that the circuitousness of Internet paths depends on the subcontinental regions where the source and destination hosts are located~\cite{6663531}. 

From a different perspective, the properties of the Internet were put in relationship with the European cities hosting its infrastructural elements~\cite{doi:10.1080/10630732.2011.578408}. In particular, the role of the cities in the European Internet was evaluated according to different metrics of centrality, ranging from simple ones, like degree centrality and weighted degree centrality, to more complex ones, such as betweenness and eigenvector centrality \cite{Newman2016}.

\subsection{IP geolocation}

Whenever the geographic aspect of the Internet is relevant, one of the first steps is to estimate the position of networking infrastructure. An IP address can be mapped to a location according to different techniques. Our study, as detailed in Section~\ref{subsec:aux-dataset}, is based on an active geolocation method, provided by RIPE IPmap~\cite{ipmap-under-the-hood}. In active geolocation, the position is estimated by probing the target IP address from a set of hosts with known location (generally called landmarks), and converting the collected latency values into distances. Then, the constraints expressed by distances and the positions of landmarks are combined to estimate the location of the target. In some cases, traceroutes are used to include topological information in the estimation process~\cite{10.1145/1177080.1177090}. These techniques have been extensively used in the past, and systems based on these principles of operation include constraint-based geolocation \cite{10.1109/TNET.2006.886332} and Spotter~\cite{5935165}. It is also possible to estimate the position of a host according to techniques that do not rely on active measurements. A first example is geolocation through reverse DNS, where the recurring structure of names adopted by operators may reveal the location of infrastructure elements. Some router names, for instance, contain city codes and similar abbreviations. In some cases, the rules needed for parsing the names are manually generated, while in other cases the most probable locations are found by interpreting reverse DNS names through machine learning methods. Notable examples include undns~\cite{10.1145/964725.633039}, DRoP~\cite{huffaker2014drop}, and RDNS~\cite{dan2018ip}. Another example is the use of crowdsourced data: users voluntarily provide the location of the IP addresses they manage to a central repository, so that locations can be subsequently retrieved by all interested users~\cite{old-openipmap}.

Besides categorizing the methods in active and non-active ones, there are other relevant properties which can be considered, among them: the resolution of the provided location estimate (country, city), if the method works better with targets belonging to the infrastructure or to end users' networks (several commercial databases are optimized for localizing the end users). 

\subsection{Topology and locality}

Africa has been the focus of several studies aimed at understanding how the topology of the Internet affects the somehow disappointing performance experienced by the  users in the region~\cite{isah2019state}.
An analysis of the Internet delay in Africa revealed that the continent is characterized by significant differences~\cite{8486024}. A few African countries have inter-country delay values that are comparable with the ones observed in Europe and North America, whereas in other countries the delay is one order of magnitude higher. In addition, some clusters of relatively well-connected countries can be identified. The main reason behind the not excellent situation of the African Internet latency was found in an excessive adoption of transit providers located in other continents, mostly in Europe and North America (i.e., a significant fraction of paths goes outside the African region even though the source and destination hosts are in Africa). 

The African Internet was studied also in terms of path locality, finding that a significant fraction of routes passes through European and North American IXPs leading to unnecessary high latency~\cite{FANOU2017117}. To this purpose, the observed RTT was compared to the minimum theoretical RTT, and significant inflation, in particular for West Africa, was found. This study, hence, is probably the closest one with respect to the one presented in this paper. However, in~\cite{FANOU2017117}, locality of paths was not the main focus of the study, which was, instead, mostly dedicated to the analysis of the topological changes that occurred during a four-year interval.
Non-local paths were found also in the Middle East, again negatively associated with some other metrics such as the RTT, and the number of hops~\cite{8845104}.

Connectivity clusters were identified also in the Latin American and Caribbean (LAC) region~\cite{10.1145/2940116.2940130}, using measurements collected by means of the Speedchecker platform. However, in this case, the geographic position of intermediate nodes was not evaluated, thus providing limited information from the point of view of path locality.  
A study about latency in China conducted by Zhuang et al. proved that a significant fraction of delay can be attributed to the excessive circuitousness of paths~\cite{ZHUANG2020107102}. In particular, consistently with~\cite{Nur:2018:GRI:3276978.3239162}, the intra-domain paths are characterized by a reduced level of circuitousness when compared to inter-domain paths. Results suggest that a sub-optimal selection of IXPs is the reason leading to many highly circuitous paths. According to Zhuang et al. this is mostly due to historical reasons, as the oldest IXPs are still used despite the presence of newly deployed, and better placed, ones.

IXP Country Jedi is the first tool that was specifically designed to study the locality of paths~\cite{aben2017ixp}, with a later variant also including the amount of user population~\cite{10.1145/3106328.3106332}. The tool, which relies on information collected by RIPE Atlas, provides a matrix-based view where the level of path locality for AS pairs in a country is visually represented. 

Path locality is also related to the geographic aspects of Internet-based transmission of copyrighted material, in case crossing of borders is not allowed~\cite{DBLP:conf/icwsm/IbosiolaSGSUT18}.

Compared to existing literature, our study provides contributions along the following directions: 

\begin{itemize}

    \item Path locality is studied from the points of view of two different metrics, one based on the address space and the other one on the served population (Section \ref{sec:locality}). By assessing the locality of paths according to these two different criteria, some characteristics of the phenomenon are better captured.
    Sections \ref{sec:results}, \ref{sec:regionscountries}, and \ref{sec:ipv6} report and discuss cases where differences between the two criteria have been observed.

    \item All the regions of the world are included in the study. This is important not only because some continents have been scarcely covered in the past, but also because the availability of results for all the regions makes possible a comparison between them. The main findings in terms of path locality are combined with other important metrics such as the RTT, and the length of IP and AS paths (Sections \ref{sec:results}). Our study relies on an extensive and rich dataset described in Section \ref{sec:dataset}. 
    
    \item A detailed analysis is provided for all the considered regions, identifying the most relevant countries, and discussing the presence of IXPs and international carriers (Section \ref{sec:regionscountries}). Some results about IPv6 are also presented (Section \ref{sec:ipv6}).  

    \item Non-local paths are used to define a dependency graph between countries (Section \ref{sec:dependency}).
    
\end{itemize}

\section{Path locality}
\label{sec:locality}

Let us call $G$ the geographic area under study, where $G$ can be a continent, a country, or an area defined according to other criteria. Let $s$ and $d$ be two hosts located in $G$, and let $(s, d) = \{s, r_1, r_2, ..., r_n, d\}$ be the path from $s$ to $d$ at the IP level, where $r_i$ is the $i$th router along the path. A path is defined as \emph{local} only if all routers $r_i$, with $i \in 1..n$, are located in $G$. Conversely, a path is non-local if at least one of the routers is not located in $G$\footnote{In practice, the data collection and processing is prone to errors which need to be handled. In Section~\ref{sec:dataset} we describe how a path is to be considered local, according to the available dataset, together with other data related issues.}.
More formally, the following function can be defined:

\begin{equation}
l(s, d) = \begin{cases}
          1 \quad &\text{if} \, \forall i , r_i \prec G\\
          0 \quad &\text{if} \, \exists i : r_i \nprec G\\
          \end{cases}
          \label{eq:onepathlocality}
\end{equation}
with $i \in 1..n$, and where the $\prec$ and $\nprec$ symbols are used to indicate inclusion, and not inclusion, within the physical boundaries of the considered area. Then, 
the path locality $L_G$ of an area $G$ can be defined as the fraction of paths that remain confined within the area: 

\begin{equation}
    L_G = \frac{\sum\limits_{s\ingeo G, d \ingeo G, s \neq d} l(s, d)}{N(N - 1)}
    \label{eq:firstdefinition}
\end{equation}
where $N$ is the total number of IP addresses belonging to the area $G$. The total number of source-destination pairs is $N(N - 1)$, since a source host does not measure the path towards itself.

Computing the locality as specified in Equation~\ref{eq:firstdefinition} is unnecessarily costly, as the number of hosts can be extremely large, especially for geographic areas like countries or continents. Thus to reduce the number of measurements, another definition can be provided by considering that the path between the source and destination hosts depends on the two subnetworks the two hosts belong to. We thus assume that all the paths from a source network to a destination network share the same locality properties, i.e. they are all local, or all non-local.
Thus, let us define $\bm{s}$ and $\bm{d}$ the source and destination networks, with $\bm{s} \ingeo G$ and $\bm{d} \ingeo G$, $s$ a single random host with $s \in \bm{s}$, and $d$ a single random host with $d \in \bm{d}$. Path locality can now be defined as 

\begin{equation}
    L'_G = \frac{\sum\limits_{\bm{s} \neq \bm{d}} l(s, d) \cdot \lvert \bm{s}\rvert \cdot \lvert \bm{d}\rvert + \sum\limits_{\bm{s} = \bm{d}} \lvert \bm{s} \rvert \cdot (\lvert \bm{d} \rvert - 1)}
    {{\sum\limits_{\bm{s} \neq \bm{d}} \lvert \bm{s}\rvert \cdot \lvert \bm{d}\rvert} + \sum\limits_{\bm{s} = \bm{d}} \lvert \bm{s} \rvert \cdot (\lvert \bm{d} \rvert - 1)}
    \label{eq:seconddefinition}
\end{equation}
To better explain Equation~\ref{eq:seconddefinition} let us consider each component on its own. The numerator shows two summations: the first considers the case of hosts belonging to different subnetworks. The second, the case of hosts belonging to the same subnetwork. For the first summation, as aforementioned, we assume all the paths from a source network to a destination network to share the same locality properties, thus we can just measure the locality of a single path, i.e. $l(s, d)$, and multiply it for the total number of paths between the subnetworks $\bm{s}$ and $\bm{d}$, to obtain the number of local paths between different networks. For the second summation, we reasonably assume that the paths between hosts of the same network are local, thus we implicitly consider $l(s, d) = 1$. We just need to calculate, for each subnetwork, the total number of paths between its hosts, which is $\lvert \bm{s} \rvert \cdot (\lvert \bm{d} \rvert - 1)$, as a host can not issue a measurement towards itself. The denominator instead just computes the total number of paths in the case that $\bm{s}$ and $\bm{d}$ are different networks (first summation), and within the same network (second summation).
Equation~\ref{eq:seconddefinition} would produce approximately the same\footnote{Some mechanisms could lead to different IP-level paths for $(s_1, d_1)$ and $(s_2, d_2)$, even when $s_1$ and $s_2$ belong to the same subnetwork $\bm{s}$, and $d_1$ and $d_2$ belong to the same subnetwork $\bm{d}$. Examples of such mechanisms include traffic engineering policies and load balancing. However, such mechanisms are adopted in current networks mostly because of technological and performance-driven reasons, so they are introduced for necessities that are orthogonal with respect to pure routing. We thus believe that almost always $l(s_1, d_1) = l(s_2, d_2)$, thus leading to $L=L'$ with very good approximation.} result of Equation~\ref{eq:firstdefinition} but with much smaller costs, as the number $M$ of subnetworks in $G$ is generally much smaller than the number of hosts $N$. The number of paths to be probed would be reduced from $N(N - 1)$ to $M(M - 1)$.

\subsection{Computing path locality in the real world}

Unfortunately, even the less expensive definition of path locality, given by Equation~\ref{eq:seconddefinition}, is far from being computable in the real world. Computing $L'$  requires the collection of $M(M-1)$ paths, from every subnetwork in $G$ to every other subnetwork in $G$. Although the total number of paths is probably manageable, no currently available measurement infrastructure is able to provide a vantage point in every subnetwork of large geographic areas, like countries or continents.
In particular, multiple issues need to be faced: i) the number of the available vantage points in $G$ is usually much smaller than the number of the subnetworks; ii) not all the measurements are successful, thus even if a source-destination pair is available in a measurement infrastructure, it can still not provide a path locality measure; iii) the collected data is subject to errors which are specific to the particular source of data, thus it has to be processed to ensure that the inaccuracies are minimized.

To cope with the first issue, i.e. the lack of vantage points, we choose to focus on ASes rather than on subnetworks. An AS is defined as a group of subnetworks controlled by a single and defined administration authority. The number of ASes in a geographic area $G$ is much smaller than the number of subnetworks, and this can help reduce the quantity of measurements needed. Thus, we group measurements by pairs of ASes.  To obtain a reasonably wide coverage in terms of ASes and geographic spread, we choose to rely on RIPE Atlas~\cite{staff2015ripe}, which is characterized by a significant presence in all the regions of the world~\cite{Faggiani2014:study}.  The coverage of the collected dataset, in terms of address space, networks, and ASes, is provided in Section~\ref{sec:dataset}. Possible limitations are instead discussed in Section~\ref{sec:discussion}. To cope with the second issue, we choose to focus only on source-destination pairs that produce successful measurements, discarding all the other pairs. The definition of successful measurements is provided in Section~\ref{sec:dataset}, which describes all the issues related to the dataset used in this work. Section~\ref{sec:dataset} also describes the data handling process that solves the third issue.

We thus define
\begin{equation}
L^{SD}_G = \frac{\sum\limits_{s \ingeo G, d \ingeo G, s \in \bm{S}, d \in \bm{D}} l(s,d)}{N^{SD}_G}
\label{eq:approxlocalitypair}
\end{equation}
as the path locality in $G$ for the pair of ASes $\bm{S}$ and $\bm{D}$, where $N^{SD}_G$ is the number of successful measurements between source-destination pairs that produce at least a result and that are located in $G$, with $s \in \bm{S}$ and $d \in \bm{D}$\footnote{It must be noted that $\bm{S}$ and $\bm{D}$ can be the same AS.}. In other words $L^{SD}_G$ is the fraction of successful measurements between $\bm{S}$ and $\bm{D}$ that remain local to the region $G$. Then, to approximate the path locality of the entire region $G$, we assume that each pair of ASes in $G$ is somehow representative of the path locality of $G$, according to their dimension. The bigger the ASes, the more representative of the locality of the whole region the pair is. We thus assign a weight to each pair of ASes, based on their dimension. More formally we define the weight of a pair of ASes $(\bm{S}, \bm{D})$ as
\begin{equation}
    w^{SD}_G = \frac{\lvert \bm{S} \rvert \cdot \lvert \bm{D} \rvert}{\sum\limits_{(\bm{A}, \bm{B}) \in \bm{\mathcal{S}}_G} \lvert \bm{A} \rvert \cdot \lvert \bm{B} \rvert}
    \label{eq:approxlocalityweight}
\end{equation}
where $\bm{\mathcal{S}}_G$ is the set of all the pairs of ASes for which there are successful measurements between source-destination pairs geolocated in $G$. Finally, we can define our approximated path locality metric for the area $G$ as
\begin{equation}
    \widehat{L_G} = \sum\limits_{(\bm{S}, \bm{D}) \in \bm{\mathcal{S}}_G} L^{SD}_G \cdot w^{SD}_G
    \label{eq:approxlocalityregion}
\end{equation}
The path locality, as computed with the equations defined above, is a number between 0 and 1. The more the locality is near to 1, the more the paths of the given geographic area are local.

As a measure of the dimension of an AS, we consider: i) the dimension of the announced address space, i.e. the total number of IP addresses that are announced by that AS via the BGP routing protocol; ii) the number of consumers that an AS serves. We thus define two locality metrics: $\widehat{L^A_G}$, which is based on the address space of ASes; $\widehat{L^C_G}$, which is based on the number of end users. The equation to compute $\widehat{L^A_G}$ and $\widehat{L^C_G}$ and is the same (Equation \ref{eq:approxlocalityregion}), the only changes are related to the dimensions of the ASes used in Equation~\ref{eq:approxlocalityweight} when computing the weights. The first metric includes all types of Internet access, including both the ASes which serve end users and other kinds of ASes such as academic, international ISPs, etc. In the second one, the ASes that have the highest weight are the ones that serve the highest number of end users. In this second case, in general, consumer ISPs have higher weight than Content Delivery Networks (CDNs) and transit providers.

The goodness of this approximation of the real path locality of a region depends on the number of the ASes of that region covered by a measurement infrastructure. To be sure to obtain a reasonable approximation, as mentioned above, in this study we use RIPE Atlas, which is the largest public Internet measurements infrastructure available, from both an AS coverage and a geographic coverage point of view~\cite{Faggiani2014:study}.

\section{Dataset}\label{sec:dataset}

\begin{table*}[t!]
    \begin{center}
        \caption{Dataset characterization; the IP coverage is calculated based on the addresses of the sources and targets involved in measurements.}
        \label{tab:datasetcharacterization}
        \resizebox{\textwidth}{!}{
        \begin{tabular}{lrrrrrrrr}
            \toprule
            & \multicolumn{4}{c}{\textbf{IPv4 Coverage}} & \multicolumn{4}{c}{\textbf{IPv6 Coverage}}\tabularnewline
            \textbf{Region} & \textbf{Traceroutes} & \textbf{Addresses} & \textbf{/24 Networks} & \textbf{ASes} & \textbf{Traceroutes} & \textbf{Addresses} & \textbf{/48 Networks} & \textbf{ASes} \tabularnewline
            \midrule
            Africa & 35\,697 & 497 & 459 & 181 & 790 & 58 & 54 & 45\tabularnewline
            Asia & 142\,489 & 3\,789 & 3\,248 & 1\,126 & 8\,259 & 600 & 498 & 238\tabularnewline
            Europe & 825\,918 & 12\,966 & 11\,248 & 2\,279 & 60\,150 & 4\,340 & 3\,899 & 998\tabularnewline
            Middle East & 41\,192 & 737 & 671 & 210 & 393 & 52 & 46 & 38\tabularnewline
            North America & 282\,510 & 3\,154 & 2\,778 & 601 & 30\,232 & 896 & 803 & 230\tabularnewline
            Oceania & 41\,501 & 548 & 503 & 134 & 2\,325 & 161 & 132 & 60\tabularnewline
            South America & 25\,537 & 715 & 658 & 234 & 1\,659 & 91 & 86 & 46\tabularnewline
            \bottomrule
        \end{tabular}
        }
    \end{center}
\end{table*}

The dataset is composed of a large number of ICMP traceroutes, collected in seven regions: Africa, Asia, Europe, Middle East, North America, Oceania, and South America. Traceroute is a network diagnostic tool commonly used to discover the IP level path between a source host and a destination host. ICMP traceroute sends multiple ICMP packets towards the destination, with increasing Time To Live (TTL) values. This is done to solicit ICMP time exceeded responses from intermediate routers along the path to identify them. The output of traceroute is an ordered sequence of IP addresses belonging to the traversed routers; i.e., the path from the source to the destination. Each step in this sequence is called hop. Traceroute is thus useful to observe routing decisions by studying how the injected packets move from a source to a destination.

\subsection{Data collection}\label{subsec:data-collection}

Measurements were collected using RIPE Atlas, a community-based Internet measurement platform composed of devices distributed worldwide, which can be instructed in performing network measurements.
Currently, RIPE Atlas is composed of more than 11\,000 devices, called probes, spread all over the world which gather more than 10\,000 measurement results per second~\cite{atlas}. The deployment of probes is denser in Europe and North America, but the large number of devices enables, in any case, an unprecedented coverage of all the regions.
In our dataset, RIPE Atlas probes are both used as sources and targets of traceroutes. 
In particular, our dataset is the union of:

\begin{enumerate}

\item All measurements performed by RIPE Atlas in the two weeks 15--28 May 2019, including both measurements performed autonomously by the platform and measurements defined by the users. 

\item A full mesh of measurements among the probes, that we performed approximately during the same time frame to further increase the total number of measurements. In the regions where probes are particularly abundant, their number was limited to 500 in our full-mesh measurements. Probes were selected to be uniformly spread across all the countries of such regions, and with the highest possible degree of AS diversity, with the same approach we adopted in~\cite{candela2019using}. For the regions with less than 500 probes, all the probes were selected.
For each pair of probes, in each region, we issued a traceroute every hour in both directions, for an entire day.
Different pairs were scheduled with a gap of $20$ seconds between each other, in order to reduce ICMP rate limiting on shared nodes~\cite{iodice2019periodic}.
\end{enumerate}

\noindent In total, we collected more than 300 million traceroutes. However, as mentioned in Section~\ref{sec:locality}, in our analysis we consider only the successful measurements, which we define as the traceroutes that are able to reach their destination. In addition, we select only one traceroute for each source-target pair. In particular, to be conservative, we consider just the traceroute reaching the destination with the lowest RTT, as it has a higher chance of being shorter, and therefore a higher chance of being local. In addition, the traceroute with the lowest RTT should be the least affected by queuing and processing delays, therefore the most reliable for comparing delays of local paths versus non-local ones. 
The final amount of traceroutes for the regions is: 36\,487 for Africa, 151\,018 for Asia, 886\,068 for Europe, 41\,585 for the Middle East, 312\,742 for North America,  43\,826 for Oceania, and 27\,196 for South America.
To better characterize our dataset, we computed the coverage at the IPv4 and IPv6 level, which is presented in Table~\ref{tab:datasetcharacterization}. The table shows, for each version of the IP protocol, the number of traceroutes, the number of unique addresses that are sources or destinations in our dataset, and the number of networks (/24 for IPv4 and /48 for IPv6) and ASes covered by these addresses (instead, the AS coverage of all the hops in our dataset is provided further). For IPv4, the coverage is good in all the regions. For IPv6, the number of measurements and the coverage is sufficient to derive some conclusions, except for a couple of regions (Africa and the Middle East), which show a low number of measurements and of covered networks/ASes.

\subsection{Data enrichment}\label{subsec:aux-dataset}

The raw traceroute results were enriched as follows.

\noindent\textbf{Geolocation.} 
The geographic locations of the probes used as sources (and sometimes as destinations) of traceroutes are provided directly by RIPE Atlas.
However, to estimate the path locality we have to know also where the intermediate hops of a traceroute are located, in particular we need a way to geolocate IP addresses belonging to the infrastructure of the Internet. In previous works, commercial or crowdsourced (i.e., manually produced) datasets have been used. However, commercial datasets are not tailored for infrastructure geolocation and have been proven not accurate~\cite{geocompare,Geo-database-comparison,candelaCCR,poese2011ip}, while the literature about the accuracy and coverage of crowdsourced datasets is not particularly abundant.

On the other hand, active geolocation has been proven to be effective in such a task~\cite{candela2019using}.
RIPE IPmap~\cite{ipmap-under-the-hood} is a geolocation platform, which provides various geolocation methodologies. One of the methodologies is based on active geolocation, where the position is calculated by means of latency measurements.
RIPE IPmap active geolocation has been reported to be 100\% accurate at the continent level, and 99.58\% at the country level~\cite{iordanou2018tracing}.
Additionally, a more recent study~\cite{candelaCCR} estimated its accuracy 80.3\% at city level (higher accuracy compared to some commercial datasets).
The study reports a median error distance of 29.03 km (more than enough accurate for our country-level analysis), and coverage of 78.5\%.

For each traceroute in our dataset, we geolocated each hop at the country level. As mentioned above, the geolocation of the probes (even when behind NAT) is directly provided by the RIPE Atlas platform, together with the public IP address they use to access the Internet. Other private IP addresses along the path are instead discarded, together with non-responding routers, as they are impossible to geolocate. On average, 67.9\% of the hops of a path have been geolocated.
Since geolocation data is incomplete and may be affected by inaccuracies, we applied the following constraints to make the dataset and the analysis more robust: 

\begin{enumerate}

\item In fiber, signals travel at approximately $2/3~c$  where $c$ is the speed of light~\cite{bovy2002analysis}. We verified that, for every geolocated router, its distance from the source is compatible with such maximum speed. If the condition does not hold, the position of the node is considered to be incorrect and the node is then reverted to the ``unlocalized'' status. 

\item To adopt a conservative approach we assume that a single router located in a different area is not enough to flag a path as non-local. To mark a path as non-local, one of the following two conditions must hold: i) at least two routers are located outside the area, ii) a router is located outside the area and it is preceded or followed by a router that cannot be located. We prefer to be conservative and consider a path as local unless possibly more than one router is located outside the considered area.

\end{enumerate}

\noindent\textbf{Peering LANs.}
If the IP address of a hop belongs to a peering LAN of an IXP, we annotate the hop with the corresponding IXP identifier and location provided by PeeringDB~\cite{peerdb}. 

\noindent\textbf{Autonomous Systems.}
We also annotated all the sources, targets, and intermediate hops with the corresponding AS number originating the prefix in which the IP at issue is contained. For RIPE Atlas probes, the AS is directly provided by the platform. For all the other hops, this step can be performed using BGP data (e.g.,~\cite{ris}). However, links between two ASes might make the IP-to-AS mapping unreliable, as every address could belong to any of the two ASes. To mitigate this problem, we adopted the methodology described in~\cite{marder2016map}. 
Overall, the dataset is composed of 4\,171 unique ASes hosting a source of measurements, and 2\,934 unique ASes hosting a target. When including also the intermediate nodes, the full dataset comprises 6\,521 unique ASes. To compute the weights in Equation~\ref{eq:approxlocalityweight}, we used two datasets. For the $\widehat{L^A_G}$ metric, for each source or destination AS in our measurements, we collected the announced prefixes, as seen by RIS~\cite{ris}, and we used them to compute the total number of IPs announced. For the $\widehat{L^C_G}$ metric, we used the APNIC's Customers per AS Measurements (ASpop) dataset~\cite{aspop}, which provides the estimated number of end users per AS.

\section{Global results}
\label{sec:results}

\begin{table}[t!]
    \begin{center}
        \caption{Total number of paths, local paths, and path locality for each world region.}
        \label{tab:pathlocality}
        \resizebox{\columnwidth}{!}{
        \begin{tabular}{lrrrr}
            \toprule
            \textbf{Region} & \textbf{Paths} & \textbf{Local paths (\%)} & $\bm{\widehat{L^A}}$ & $\bm{\widehat{L^C}}$ \tabularnewline
            \midrule
            Africa & 35\,697 & 20\,755 (58.1\%) & 0.638 & 0.327\tabularnewline
            Asia & 142\,489 & 92\,018 (64.6\%) & 0.811 & 0.772 \tabularnewline
            Europe & 825\,918 & 819\,891 (99.3\%) & 0.990 & 0.982 \tabularnewline
            Middle East & 41\,192 & 20\,381 (49.8\%) & 0.417 & 0.420 \tabularnewline
            North America & 282\,510 & 276\,204 (97.8\%) & 0.962 & 0.988 \tabularnewline
            Oceania & 41\,501 & 39\,909 (96.2\%) & 0.999 & 0.996 \tabularnewline
            South America & 25\,537 & 16\,275 (63.7\%) & 0.809 & 0.601 \tabularnewline
            \bottomrule
        \end{tabular}
        }
    \end{center}
\end{table}

We present the main results at the region level.
Table~\ref{tab:pathlocality} shows the number of paths collected for each region, the number (and percentage) of paths that remain local to the region, and the path locality values, for IPv4. 

We first consider the path locality computed with the address spaces, $\widehat{L^A}$. From a first analysis, we can divide the world into three groups of regions according to their path locality values: a group of very well connected areas composed of Europe, North America, and Oceania, which shows values of path locality around 0.96--1; a group with intermediate path locality values, around 0.8, composed by Asia and South America; and a group of less connected areas composed by Africa and the Middle East, with path locality that ranges from approximately 0.4 in the Middle East to 0.64 in Africa.

The path locality values, overall, seem to follow the economic and technological characteristics of the three groups of regions. Europe, North America, and Oceania are, on average, high-income regions characterized by a high level in the Information and Communications Technology (ICT) domain, according to the ICT Development Index (IDI) published by ITU~\cite{international2017measuring}. IDI is a composite benchmarking index based on a number of sub-indicators concerning access, use, and skills in ICT. 
For these regions, the value of $\widehat{L^A}$ is approximately equal to the raw percentage of local paths. Asia and South America include large areas characterized by rapid economic and technological growth, but also some areas with lower levels of Gross National Income (GNI) and IDI. This could explain why these two regions are not entirely self-containing from a path locality point of view. Africa and the Middle East include a number of low- and middle-income countries, and they are also characterized by generally lower IDI values. The low percentage of path locality in these regions could thus be explained by the reduced general performance in the ICT domain. In addition, in Africa and the Middle East, several countries are characterized by a non-idyllic situation from the point of view of conflicts and political stability, as summarized by their Global Conflict Risk Index~\cite{gcri}. For the latter four regions, except the Middle East, the values of $\widehat{L^A}$ are higher than the percentage of local paths. This means that the AS pairs that show the highest weights are connected by local paths. On the contrary, the Middle East shows an opposite situation, with a $\widehat{L^A}$ value smaller than the percentage of local paths.

We now consider the path locality computed with the number of end users per AS, $\widehat{L^C}$. We can observe that the group composed of Europe, North America, and Oceania obtains the same results. This means that, in these regions, the Internet paths are always local, even the ones that connect the end users. Also Asia and the Middle East show almost unchanged path locality for end users connectivity. Africa and South America instead show a significant drop in path locality when the focus is on the paths that connect end users. Africa is characterized by an extremely low value of $\widehat{L^C}$ of 0.327, which means that a large fraction of paths that connect end users flow outside the region. In general, we observe that for regions characterized by lower GNI and IDI, $\widehat{L^C}$ is less or equal to $\widehat{L^A}$. This means that the ASes with a high weight in $\widehat{L^C}$ (i.e., the ASes that serve end users) seem to struggle more in keeping their traffic local, than other ASes such CDNs, transit providers, etc, which have a high weight in $\widehat{L^A}$.

\subsection{Path locality towards content targets}

\begin{table}[t!]
    \begin{center}
        \caption{Total number of paths, local paths, and path locality for each world region, towards content targets (i.e., Google, YouTube, Netflix, Akamai, Amazon, Fastly, Cloudflare, Microsoft, Facebook, Twitter, LinkedIn)}
        \label{tab:reg-countries-content}
        \resizebox{\columnwidth}{!}{
        \begin{tabular}{lrrrr}
            \toprule
            \textbf{Region} & \textbf{Paths} & \textbf{Local paths (\%)} & $\bm{\widehat{L^A}}$ & $\bm{\widehat{L^C}}$ \tabularnewline
            \midrule
            Asia & 6\,814 & 5\,306 (77.9\%) & 0.926 & 0.995 \tabularnewline
            Europe & 26\,211 & 25\,526 (97.4\%) & 0.988 & 0.998 \tabularnewline
            North America & 12\,790 & 11\,834 (92.5\%) & 0.937 & 0.972 \tabularnewline
            Oceania & 477 & 473 (99.2\%) & 1 & 1 \tabularnewline
            South America & 421 & 383 (91.0\%) & 0.905 & 0.817 \tabularnewline
            \bottomrule
        \end{tabular}
        }
    \end{center}
\end{table}

Content providers are a crucial part of the Internet, as they serve a considerable amount of the data end users are interested in: videos, images, social networking, etc. 
Bringing content as close as possible to end users is one of the challenges of the modern Internet. 
In this section, we analyze the locality of paths that are related to content providers. For each region, we restricted the paths to the ones that reach only the addresses of content providers. In particular, from the traceroutes with source and destination in a given region, we choose the ones with the destination belonging to a content provider's network. The content providers we considered are Google, YouTube, Netflix, Akamai, Amazon, Fastly, Cloudflare, Microsoft, Facebook, Twitter, and LinkedIn. Results are shown in Table~\ref{tab:reg-countries-content}. We excluded Africa and the Middle East from the results, as the number of paths for these regions was too small to derive conclusions: 52 and 10, respectively. As can be observed, the number of paths towards content varies substantially among regions. The percentage of local paths is extremely high, as well as the path locality values which are always close to 1. The only exception is Asia, which shows just 77.9\% of local paths, but maintains high values for $\widehat{L^A}$ and $\widehat{L^C}$. This means that the ASes with the highest weights, according to both metrics, manage to maintain local paths toward content. In conclusion, the content infrastructure seems to be well connected, even in regions that do not show extremely high values of path locality.

\subsection{Impact on other metrics}

\begin{table*}[t!]
  \begin{center}
    \caption{Properties of local and non-local paths.}
    \label{tab:metrics}
    \begin{tabular}{l  r r r r r r r r} 
    \toprule
      \textbf{Region} & \multicolumn{2}{c}{\textbf{IP path length (hops)}} & \multicolumn{2}{c}{\textbf{AS path length (hops)}} & \multicolumn{2}{c}{\textbf{RTT (ms)}} & \multicolumn{2}{c}{\textbf{Path length (km)}} \tabularnewline
      & Local & Non-local & Local & Non-local & Local & Non-local & Local & Non-local \tabularnewline
      \midrule
      Africa & 13.8 & 18.9 & 4.7 & 5.5 & 70.1 & 257.6 & 5\,752 & 22\,538 \tabularnewline
      Asia & 13.5 & 17.7 & 5.0 & 6.1 & 106.9 & 267.4 & 10\,168 & 27\,700 \tabularnewline
      Europe & 11.9 & 14.4 & 4.8 & 5.5 & 31.8 & 71.3 & 3\,228 & 15\,459 \tabularnewline
      Middle East & 14.7 & 20.7 & 3.7 & 5.7 & 87.8 & 171.2 & 1\,670 & 11\,281 \tabularnewline
      North America & 14.1 & 14.5 & 4.6 & 4.8 & 56.4 & 95.6 & 4\,168 & 18\,634 \tabularnewline
      Oceania & 13.4 & 20.0 & 4.5 & 6.3 & 42.0 & 285.0 & 5\,275 & 28\,997 \tabularnewline
      South America & 14.3 & 19.3 & 4.7 & 5.9 & 71.7 & 216.8 & 4\,323 & 16\,897 \tabularnewline
      \bottomrule
    \end{tabular}
  \end{center}
\end{table*}

Here, we analyze the path locality of all the regions together with other important metrics related to the geographic side of the Internet. The considered metrics are the length of both IP and AS paths, the RTT, and the path length in kilometers. To compute the length of a path in kilometers, we positioned each hop of a path in the centroid of the country where it was geolocated, discarding the non-geolocated ones. Then, we calculated the distance between the centroids of each pair of consecutive hops, and summed all the distances. While this is not to be considered an accurate measure of the actual geographic length of a path, we believe that it can give a hint about the circuitousness of local and non-local paths. Table~\ref{tab:metrics} shows the average values of the aforementioned metrics for local paths and non-local ones. As expected, the length of paths at the IP and AS level is always larger for non-local paths compared to local ones. The same applies to the RTT, but in this case, the difference between local and non-local paths is much larger. On average, for non-local paths, the IP path length and the AS path length increase 31\% and 25\%, respectively, where the RTT increases $\sim$210\%. This is due to the geographic extension of the segments used to exit from a region and then to come back, as highlighted by the increase of the geographic path length, which is on average $\sim$350\%. Such segments thus introduce a significant amount of additional delay, but they do not add much to the IP and AS path length as they generally belong to just a few ASes. Obviously, the above considerations apply to the aggregated values and it is possible that a non-local path can be better than a local one. However, the benefits of keeping the traffic local, in terms of latency, are generally quite significant.

\section{Regions and countries}
\label{sec:regionscountries}

\begin{table*}[t!]
  \footnotesize
  \centering
  \caption{Path locality of regions.\label{tab:reg-countries}}
  \subfloat[Africa]{%
    \label{subtab:africa}
    \begin{tabular}{|l r r r|}
        \hline
        \multicolumn{4}{|c|}{\textbf{Local paths}}\\
        Passing by & \# paths & $\widehat{L^A}$& $\widehat{L^C}$\\
        \hline
        All & 20\,755 & 0.638 & 0.327\\ 
        ZA & 17\,933 & 0.407 & 0.121\\
        KE & 4\,757 & 0.052 & 0.099\\
        TZ & 3\,084 & 0.002 & 0.002\\
        \hline
        \multicolumn{4}{|c|}{\textbf{Non local paths}}\\
        Passing by & \# paths & $\widehat{NL^A}$& $\widehat{NL^C}$\\
        \hline
        All & 14\,942 & 0.362 & 0.673\\
        GB & 10\,259 & 0.165 & 0.219\\
        FR & 5\,227 & 0.222 & 0.382\\
        PT & 3\,580 & 0.024 & 0.057\\
        ES & 2\,801 & 0.056 & 0.160\\
        IT & 1\,807 & 0.116 & 0.251\\
        \hline
    \end{tabular}%
  }
  \hspace{1mm}
  \subfloat[Asia]{%
  \label{subtab:asia}
    \begin{tabular}{|l r r r|}
        \hline
        \multicolumn{4}{|c|}{\textbf{Local paths}}\\
        Passing by & \# paths & $\widehat{L^A}$& $\widehat{L^C}$\\
        \hline
        All & 92\,018 & 0.811 & 0.772\\
        SG & 36\,445 & 0.096 & 0.034\\
        RU & 23\,225 & 0.018 & 0.032\\
        JP & 22\,275 & 0.352 & 0.150\\
        \hline
        \multicolumn{4}{|c|}{\textbf{Non local paths}}\\
        Passing by & \# paths & $\widehat{NL^A}$& $\widehat{NL^C}$\\
        \hline
        All & 50\,471 & 0.189 & 0.228\\
        DE & 30\,674 & 0.049 & 0.037 \\
        US & 13\,827 & 0.112 & 0.108\\
        GB & 10\,047 & 0.026 & 0.020\\
        IT & 7\,764 & 0.005 & 0.004\\
        NL & 7\,707 & 0.011 & 0.011\\
        \hline
    \end{tabular}%
  }
    \hspace{1mm}
    \subfloat[Europe]{%
    \label{subtab:europe}
    \begin{tabular}{|l r r r|}
        \hline
        \multicolumn{4}{|c|}{\textbf{Local paths}}\\
        Passing by & \# paths & $\widehat{L^A}$& $\widehat{L^C}$\\
        \hline
        All & 819\,891 & 0.990 & 0.982\\
        DE & 462\,900 & 0.447 & 0.332\\
        GB & 199\,453 & 0.358 & 0.505\\
        NL & 178\,934 & 0.191 & 0.097\\
        \hline
        \multicolumn{4}{|c|}{\textbf{Non local paths}}\\
        Passing by & \# paths & $\widehat{NL^A}$& $\widehat{NL^C}$\\
        \hline
        All & 6\,027 & 0.010 & 0.018\\
        RU & 2\,900 & 0.004 & 0.014\\
        US & 2\,002 & 0.004 & 0.001\\
        TR & 958 & 0.002 & 0.002\\
        JP & 350 & $<$0.001 & $<$0.001 \\
        HK & 87 & $<$0.001 & $<$0.001\\
        \hline
    \end{tabular}%
  }
    \hspace{1mm}
    \subfloat[Middle East]{%
    \label{subtab:middle}
    \begin{tabular}{|l r r r|}
        \hline
        \multicolumn{4}{|c|}{\textbf{Local paths}}\\
        Passing by & \# paths & $\widehat{L^A}$& $\widehat{L^C}$\\
        \hline
        All & 20\,381 & 0.417 & 0.420\\
        IR & 13\,314 & 0.050 & 0.019\\
        TR & 2\,104 & 0.047 & 0.066\\
        AE & 1\,758 & 0.107 & 0.133\\
        \hline
        \multicolumn{4}{|c|}{\textbf{Non local paths}}\\
        Passing by & \# paths & $\widehat{NL^A}$& $\widehat{NL^C}$\\
        \hline
        All & 20\,811 & 0.583 & 0.580\\
        DE & 14\,350 & 0.340 & 0.310\\
        GB & 4\,313 & 0.136 & 0.066\\
        IT & 3\,625 & 0.088 & 0.141\\
        RU & 2\,940 & 0.028 & 0.005\\
        FR & 2\,900 & 0.109 & 0.133\\
        \hline
    \end{tabular}%
  }
  \hspace{1mm}
    \subfloat[North America]{%
    \label{subtab:namerica}
    \begin{tabular}{|l r r r|}
        \hline
        \multicolumn{4}{|c|}{\textbf{Local paths}}\\
        Passing by & \# paths & $\widehat{L^A}$& $\widehat{L^C}$\\
        \hline
        All & 276\,204 & 0.962 & 0.988\\
        US & 248\,322 & 0.904 & 0.921\\
        CA & 99\,405 & 0.128 & 0.137\\
        MX & 9\,968 & 0.013 & 0.115\\
        \hline
        \multicolumn{4}{|c|}{\textbf{Non local paths}}\\
        Passing by & \# paths & $\widehat{NL^A}$& $\widehat{NL^C}$\\
        \hline
        All & 6\,306 & 0.038 & 0.012\\
        DE & 1\,467 & 0.002 & 0.009\\
        GB & 1\,300 & 0.024 & 0.001\\
        CO & 938 & 0.005 &  $\ll$0.001\\
        JP & 711 & $<$0.001 & $\ll$0.001\\
        AT & 553 & $<$0.001 & 0.001\\
        \hline
  \end{tabular}%
  }
  \hspace{1mm}
    \subfloat[Oceania]{%
    \label{subtab:oceania}
    \begin{tabular}{|l r r r|}
        \hline
        \multicolumn{4}{|c|}{\textbf{Local paths}}\\
        Passing by & \# paths & $\widehat{L^A}$& $\widehat{L^C}$\\
        \hline
        All & 39\,909 & 0.999 & 0.996\\
        AU & 34\,130 & 0.977 & 0.953\\
        NZ & 20\,310 & 0.054 & 0.212\\
        FJ & 484 & $\ll$0.001 & 0.001\\
        \hline
        \multicolumn{4}{|c|}{\textbf{Non local paths}}\\
        Passing by & \# paths & $\widehat{NL^A}$& $\widehat{NL^C}$\\
        \hline
        All & 1\,592 & 0.001 & 0.004\\
        US & 1\,075 & $<$0.001 & 0.003\\
        JP & 375 & $<$0.001 & 0.001\\
        HK & 357 & $\ll$0.001 & $<$0.001\\
        SG & 297 & $\ll$0.001 & $<$0.001\\
        MY & 13 & $\lll$0.001 & $\lll$0.001\\
        \hline
    \end{tabular}%
  }
    \hspace{1mm}
    \subfloat[South America]{%
    \label{subtab:samerica}
    \begin{tabular}{|l r r r|}
        \hline
        \multicolumn{4}{|c|}{\textbf{Local paths}}\\
        Passing by & \# paths & $\widehat{L^A}$& $\widehat{L^C}$\\
        \hline
        All & 16\,275 & 0.809 & 0.601\\
        BR & 7\,357 & 0.628 & 0.286\\
        AR & 5\,575 & 0.096 & 0.061 \\
        UY & 2\,876 & 0.050 & 0.025\\
        \hline
        \multicolumn{4}{|c|}{\textbf{Non local paths}}\\
        Passing by & \# paths & $\widehat{NL^A}$& $\widehat{NL^C}$\\
        \hline
        All & 9\,262 & 0.191 & 0.399\\
        US & 9\,238 & 0.157 & 0.386\\
        BQ & 424 & 0.002 & 0\\
        GB & 68 & 0.021 & 0.001\\
        DE & 43 & 0.052 & 0.043\\
        FR & 40 & 0.008 & 0.032\\
        \hline
    \end{tabular}%
  }
\end{table*}

In this section, we describe the geographic and topological properties of local and non-local paths, for all the seven world regions. Table~\ref{tab:reg-countries}  shows the three countries traversed by the highest number of local paths, along with the values of path locality expressed by $\widehat{L^A}$ and $\widehat{L^C}$. The path locality values for the three countries are computed by considering a modified $l(s, d)$ function in Equation~\ref{eq:onepathlocality}. In particular, $l(s, d)$ is 1 if a path is local to the region and traverses a specific country, 0 otherwise. For non-local paths we use two path non-locality metrics $\widehat{NL^A}$ and $\widehat{NL^C}$, which are defined just as $\widehat{L^A}$ and $\widehat{L^C}$, but using a function $nl(s, d)$. The latter function is similar to $l(s, d)$, but with a value equal to 1 when a path is non-local and passes through the considered country, 0 otherwise. The top five countries, traversed by the highest number of non-local paths, are also shown in Table~\ref{tab:reg-countries}. Countries are identified using ISO 3166-1 alpha-2 codes. Note that the values shown in Table~\ref{tab:reg-countries} are obtained using all the paths having source and destination in the same region, but not necessarily in the same country. Thus, the values of Table~\ref{tab:reg-countries} are not to be considered as representative of intra-country locality. Instead, in Section~\ref{sec:dependency}, we show results obtained using only intra-country measurements, to highlight dependencies among countries. 
Table~\ref{tab:topology} shows information about the topological properties: the number of local and non-local paths that flow through an IXP, through a Tier-1 AS, and through providers that are neither IXPs nor Tier-1 ASes. Some paths can traverse both an IXP and a Tier-1 AS, thus the sum of the three percentages can be slightly higher than 100\%.

\begin{table}[t!]
    \begin{center}
        \caption{Number of paths passing via IXPs, Tier-1 ASes, and other facilities (not IXP nor Tier-1 ASes). It must be noticed that a percentage of paths for each region can traverse both an IXP and a Tier-1 AS, thus the sum of the three percentages can be slightly higher than 100\%.} 
        \label{tab:topology}
        \resizebox{\columnwidth}{!}{
        \begin{tabular}{lrrrr} 
            \toprule
            \multicolumn{5}{c}{\textbf{Local paths}} \tabularnewline
            \textbf{Region} & \textbf{Total} & \textbf{Via IXPs (\%)} & \textbf{Via T1 (\%)} & \textbf{Via Other (\%)} \tabularnewline 
            \midrule
            Africa & 20\,755 & 15\,280 (73.6\%) & 1\,952 (9.4\%) & 4\,241 (20.4\%) \tabularnewline 
            Asia & 92\,018 & 31\,645 (34.4\%) & 23\,228 (25.2\%) & 37\,989 (41.3\%) \tabularnewline 
            Europe & 819\,891 & 356\,462 (43.5\%) & 289\,066 (35.6\%) & 189\,880 (23.2\%) \tabularnewline 
            Middle East & 20\,381 & 1\,093 (5.4\%) & 3\,400 (16.7\%) & 15\,916 (78.1\%) \tabularnewline 
            North America & 276\,204 & 56\,632 (20.5\%) & 133\,854 (48.5\%) & 90\,657 (32.8\%) \tabularnewline 
            Oceania & 39\,909 & 21\,080 (52.8\%) & 1\,006 (2.5\%) & 18\,056 (45.2\%) \tabularnewline 
            South America & 16\,275 & 4\,400 (27.0\%) & 7\,849 (48.2\%) & 4\,134 (25.4\%) \tabularnewline 
            \midrule
            \multicolumn{5}{c}{\textbf{Non-local paths}} \tabularnewline
            \textbf{Region} & \textbf{Total} & \textbf{Via IXPs (\%)} & \textbf{Via T1 (\%)} & \textbf{Via Other (\%)} \tabularnewline 
            \midrule
            Africa & 14\,942 & 3\,468 (23.2\%) & 9\,544 (63.9\%) & 2\,571 (17.2\%) \tabularnewline 
            Asia & 50\,471 & 19\,391 (38.4\%) & 23\,485 (46.5\%) & 8\,338 (16.5\%) \tabularnewline 
            Europe & 6\,027 & 2\,063 (34.2\%) & 3\,069 (50.9\%) & 1\,054 (17.5\%) \tabularnewline 
            Middle East & 20\,811 & 4\,665 (22.4\%) & 13\,135 (63.1\%) & 3\,168 (15.2\%) \tabularnewline 
            North America & 6\,306 & 721 (11.4\%) & 4\,482 (71.1\%) & 948 (15.0\%) \tabularnewline 
            Oceania & 1\,592 & 298 (18.7\%) & 951 (59.7\%) & 422 (26.5\%) \tabularnewline 
            South America & 9\,262 & 513 (5.5\%) & 8\,166 (88.2\%) & 603 (6.5\%) \tabularnewline 
            \bottomrule
        \end{tabular}
        }
    \end{center}
\end{table}

\subsection{Africa}

In Africa, the $\widehat{L^A}$ and $\widehat{L^C}$ values are quite different (Table~\ref{subtab:africa}). In particular, the $\widehat{L^C}$ values do not reflect the raw number of paths, especially for South Africa. This means that the ASes that serve end users struggle to keep their paths local. Almost 74\% of local paths traverse an IXP, and the presence of international Internet carriers is marginal, approximately 9\% (Table~\ref{tab:topology}).
The top five countries that are outside of the region are all European. This could be explained by the relative proximity of these countries to Africa and by the presence of submarine cables~\cite{submarinecables}. Also for non-local paths, the values of $\widehat{NL^A}$ and $\widehat{NL^C}$ often do not reflect the raw number of paths. For example, Italy has a higher value than Portugal and Spain, with significantly fewer paths. France shows the highest path non-locality values, with a relatively low number of paths traversing it. This means that there are source and destination ASes, with large address spaces and that serve many end users, which use paths that traverse France. The presence of IXPs is lower, with 23\% of non-local paths traversing an IXP, in general international ones. 
Only a few non-local paths traverse a local IXP, and this suggests how these are able to maintain traffic confined inside Africa. In non-local paths, the presence of Tier-1 ASes is prevalent, with 64\% traversing one of them. 
Many local ISPs are traversed as well by non-local paths.
This could indicate that some areas of Africa are still lacking infrastructures or peering agreements that would allow keeping more traffic local.

\subsection{Asia}

 Japan shows the highest path locality values for both metrics, even having the smallest number of traversing paths in the top three countries (Table~\ref{subtab:asia}). This means that the source-destination AS pairs that produce paths traversing Japan have a higher weight in terms of address space and served consumers. Compared to Africa, the presence of local IXPs is less relevant, but still significant, with approximately 34\% of local paths traversing at least one IXP (Table~\ref{tab:topology}).
The presence of Tier-1 ASes is significant, with 25\% of local paths traversing them, as also the presence of other providers (41\%). 
The Asian non-local paths flow mainly through Europe and North America. The most relevant country, per weight, is the USA. For non-local paths there is a strong presence of IXPs, traversed by 38\% of the paths, and of Tier-1 ASes, traversed by 47\% of the paths. The remaining 17\% of non-local paths do not traverse any of the two. The IXPs traversed by non-local paths are mainly located in Europe. 
The paths that traverse the USA are instead characterized by a negligible presence of IXPs.

\subsection{Europe}

European paths are almost always local (Table~\ref{subtab:europe}). From a topological perspective, 44\% of the local paths traverse an IXP (Table~\ref{tab:topology}). All European countries have local IXPs traversed by a non-negligible portion of paths, but three major IXPs are traversed by a significant number of paths.
Also the presence of Tier-1 ASes in local paths is significant, with 36\% of paths passing through them.
This indicates that these ASes have a pervasive presence in Europe for the routing of local traffic. 
Other providers serve a significant portion of the local paths (23\%) without the use of IXPs and Tier-1 providers.
In European local paths, we can also observe a large number of paths traversing the G\'eant network, which is the European network of academic networks. This indicates that in Europe the RIPE Atlas platform is also able to cover several academic networks, and to capture the locality of their traffic. Non-local paths account for less than 1\% of the total paths. Approximately 34\% of these paths traverse an IXP, mainly European ones, while 51\% traverse a Tier-1 AS.

\subsection{The Middle East}

The Middle East is the region with the smallest value of path locality (Table~\ref{subtab:middle}), for both $\widehat{L^A}$ and $\widehat{L^C}$ metrics, with values of 0.417 and 0.420, respectively. Most of the local paths traverse one of Iran, Turkey, or United Arab Emirates (UAE). The paths that traverse the UAE account for the highest amount of locality for both metrics, even if they are less in number than the ones of the other two countries as the source-destination pairs that produce these paths are heavier in terms of address space, and number of consumers. The presence of IXPs in local paths is minimal, with approximately 5\% of the local paths  traversing an IXP (Table~\ref{tab:topology}). 
The presence of Tier-1 ASes in local paths is higher if compared to IXPs; however, it accounts for only 17\%. The remaining 78\% of the local paths are instead traversing other, mostly local, operators. 
The non-local paths of the Middle East flow mainly through Europe and Russia. In particular, the top five countries involved in non-local paths include Germany and United Kingdom, which seem to be Internet hubs for nearby regions, including the Middle East. The presence of IXPs is higher than in local paths, with approximately 22\% of non-local paths flowing through IXPs. 
The vast majority of them traverse a European IXP, again showing the attractive force of Europe towards the Middle East, and indicating that the Middle East region is currently lacking sufficient local facilities to keep the traffic local. The presence of Tier-1 ASes in non-local paths is high, with 63\% of non-local paths traversing a Tier-1 AS.
It is also worth noting that some paths from academic networks in Israel 
are routed via the United Kingdom and the G\'eant network. In conclusion, the Middle East shows a great dependency on Europe for its non-local paths that mainly flow through European countries.

\subsection{North America}

Table~\ref{subtab:namerica} shows that North America is one of the regions with the highest values of path locality, with 0.962 and 0.988 for $\widehat{L^A}$ and $\widehat{L^C}$, respectively. As expected, USA is the most traversed country, and this makes it a sort of hub for North America. The local paths crossing an IXP are 21\% of the total (Table~\ref{tab:topology}), which is half the value of Europe, making IXPs more marginal for obtaining path locality in North America.
The presence of Tier-1 ASes in local paths is the highest among the seven regions, with 49\% of local paths traversing a Tier-1 network. A relevant presence of other providers is observed (33\% of local paths).
The non-local paths account for just 2\% of the total, with a $\widehat{NL^A}$ value of 0.038 and a $\widehat{NL^C}$ value of 0.012. IXPs and Tier-1 are traversed by 11\% and 71\% of the paths, respectively.
The traversed IXPs by non-local paths are almost always local.

\subsection{Oceania}

As shown in Table~\ref{subtab:oceania}, almost all the paths of Oceania are local, with very high locality values. Almost all paths traverse either Australia or New Zealand. Table~\ref{tab:topology} shows that the presence of IXPs in the region is very high, with approximately 53\% of local paths traversing an IXP.
The presence of Tier-1 ASes is negligible, with just 3\% of the local paths traversing one of them. Instead, a significant portion of local paths (45\%) is routed without the use of IXPs and Tier-1 providers. 
The non-local paths of Oceania account for extremely low values for both metrics, and for this reason will not be further discussed.

\subsection{South America}

In South America, the path locality values are quite different for the two metrics: $\widehat{L^A}$ is 0.809, and $\widehat{L^C}$ is 0.601 (Table~\ref{subtab:samerica}). The second reflects the actual proportions of local and non-local paths, while the first is higher. This means that the weight of the AS pairs that produce local paths in $\widehat{L^A}$ is much higher than that of the AS pairs that produce non-local paths. The percentage of local paths that traverse an IXP is 27\% (Table~\ref{tab:topology}). 
The presence of Tier-1 ASes in local paths is quite high (48\%). As in North America, Tier-1 ASes are used to route most of the local traffic. The remaining 25\% of the paths are routed without the use of IXPs or Tier-1 providers.
Almost all non-local paths are routed via the USA, which seems to be a hub for non-local traffic of South America. The presence of IXPs in non-local paths is minimal, and almost all of them  are in the USA. The rest of the paths mainly traverse a Tier-1 network.

\section{IPv6 path locality}
\label{sec:ipv6}

\begin{table}[t!]
    \begin{center}
        \caption{IPv6 Total number of paths, local paths, and path locality for each world region.}
        \label{tab:pathlocalityv6}
        \resizebox{\columnwidth}{!}{
        \begin{tabular}{lrrrr}
            \toprule
            \textbf{Region} & \textbf{Paths} & \textbf{Local paths (\%)} & $\bm{\widehat{L^A}}$ & $\bm{\widehat{L^C}}$  \tabularnewline
            \midrule
            Asia & 8\,529 & 5\,439 (63.8\%) & 0.488 & 0.886 \tabularnewline
            Europe & 60\,150 & 59\,810 (99.4\%) & 1 & 0.995 \tabularnewline
            North America & 30\,232 & 29\,941 (99.0\%) & 0.990 & 1 \tabularnewline
            Oceania & 2\,325 & 2\,126 (91.4\%) & 0.689 & 0.983 \tabularnewline
            South America & 1\,659 & 1\,233 (74.3\%) & 0.998 & 0.908 \tabularnewline
            \bottomrule
        \end{tabular}
        }
    \end{center}
\end{table}

Table~\ref{tab:pathlocalityv6} shows the path locality for IPv6 measurements only. To ensure statistical validity, we show results for the world regions that have at least 1\,000 IPv6 paths. We thus excluded Africa and the Middle East that have just 790 and 393 IPv6 paths, respectively. As can be seen from Table~\ref{tab:pathlocalityv6}, IPv6 shows some similarities with IPv4, but also some differences. As in IPv4, Europe and North America show an almost total locality of paths, for both $\widehat{L^A}$ and $\widehat{L^C}$. South America shows very high values too, differently from what happens for IPv4. It must be however noted that the percentage of paths that are local is 74.3\%, meaning that for both metrics there are some light weight pairs of ASes that still produce non-local paths. On the contrary, in Oceania the percentage of local paths is approximately 90\%, but $\widehat{L^A}$ is quite low. This means that some heavy weight pairs of ASes show a high amount of non-local paths. Asia shows a low $\widehat{L^A}$, and a higher $\widehat{L^C}$, but still it is not able to reach optimal locality levels. In the following, we analyze each region in detail.

\begin{figure*} 
\centering
\begin{tikzpicture}[      
        every node/.style={anchor=south west,inner sep=0pt},
        x=1mm, y=1mm,
      ]   
     \node (fig1) at (0,0)
       {\includegraphics[width=\textwidth]{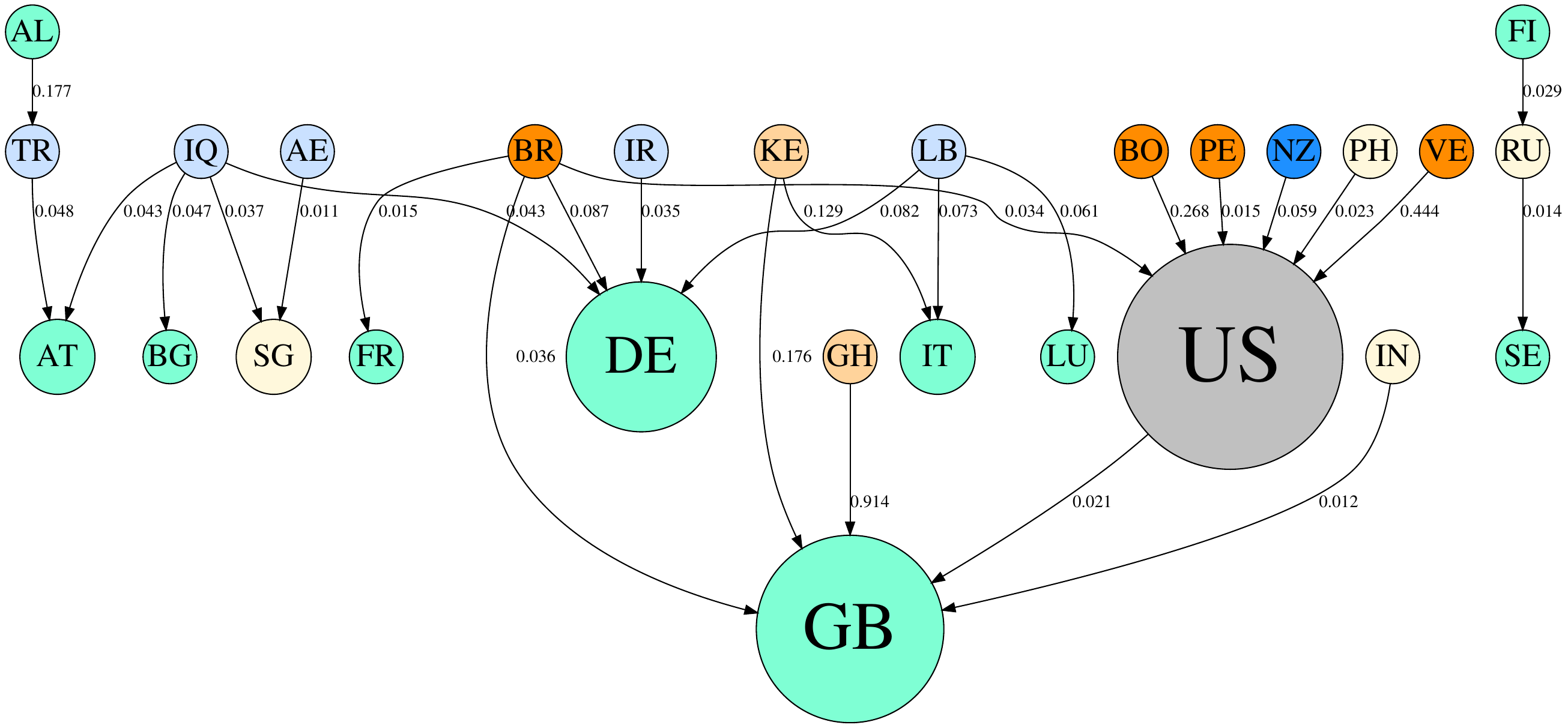}};
     \node (fig2) at (2,3)
       {\includegraphics[scale=0.21,trim=850 460 440 230,clip]{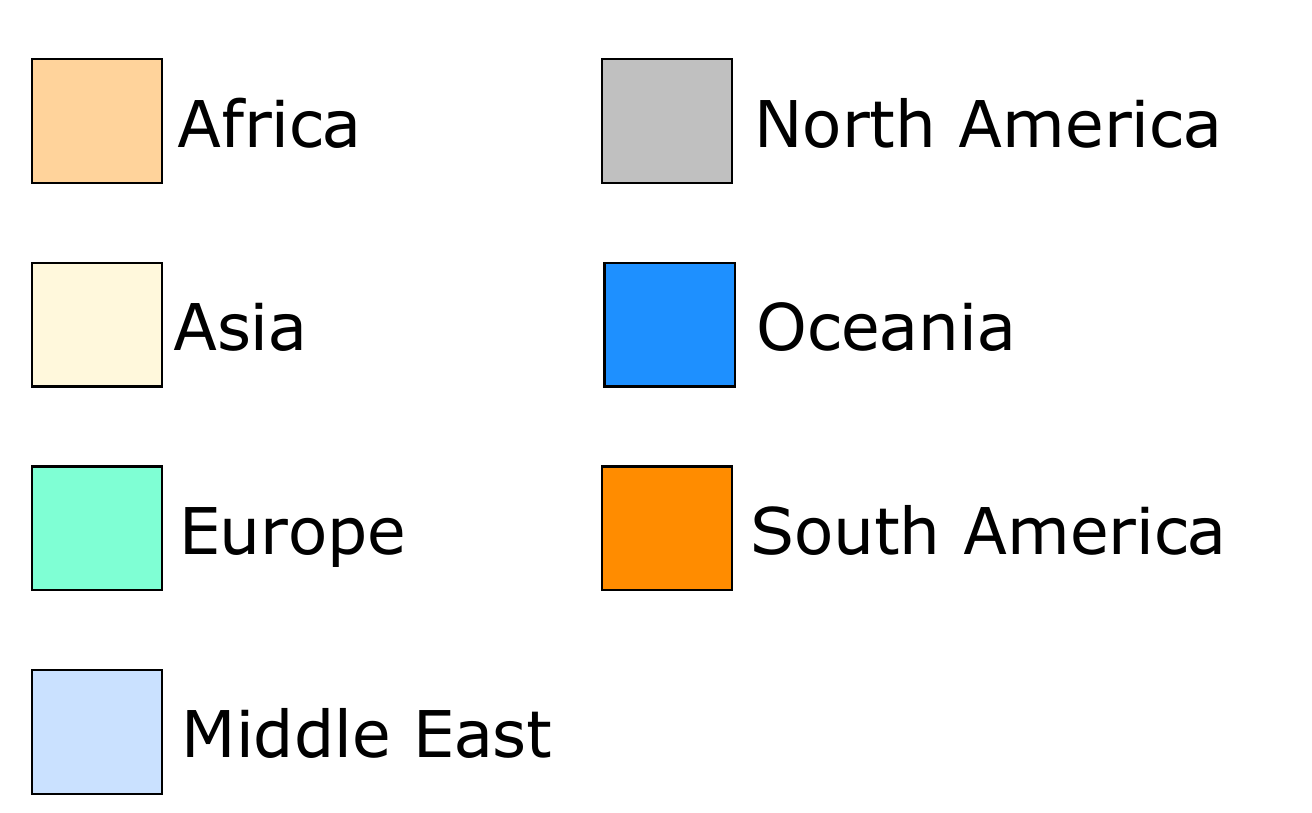}};  
\end{tikzpicture}
	\caption{Dependency graph among countries of different regions, computed on both IPv6 and IPv4 results; only countries with at least 20 source-destination pairs are considered; only edges with $\widehat{NL^A} \geq$ 0.01 are included; the size of a node $v$ is proportional to $deg^{-}_v$.}
	\label{fig:graph1}
\end{figure*}

In Asia, the most traversed countries by local paths are Singapore and Japan, and almost half of the IPv6 local paths traverse an IXP.
The most traversed countries by IPv6 non-local paths are Germany, France, USA, Sweden, and the Netherlands, with almost 2\,200 out of 3\,000 (73\%) non-local paths crossing an IXP, thus there is a considerable presence of IXPs in non-local traffic. 
European countries appear to be used for routing traffic of local operators in Russia, Kazakhstan, Armenia, and other west Asian countries, USA is instead used for routing east Asian traffic.

Europe shows some small differences with respect to IPv4. The most traversed countries by local paths are Germany, the Netherlands, and Austria. The portion of paths passing through an IXP is more than half of the total, approximately 35\,000. 
We do not analyze the European IPv6 non-local paths, as their number is so small to make them irrelevant, from a locality perspective.

The most traversed countries by North American IPv6 local paths are USA and Canada. Approximately 12\,000 local paths traverse an IXP.
North American non-local paths account for just 1\%, thus we will not analyze them.

In Oceania, as in IPv4, almost all local paths are traversing one of Australia and New Zealand. Approximately 1\,400 out of 2\,300 (61\%) local paths traverse an IXP. 
In general, IPv6 local paths show very similar behavior to IPv4 local paths. The IPv6 non-local paths are just 199, and 128 of them pass through the USA. However, they account for 0.280 of $\widehat{NL^A}$, meaning that some AS pairs with a very high number of addresses produce paths that traverse the USA.

As highlighted above, the percentage of South American IPv6 local paths is just 74.3\%, but the path locality values are close to 1 for both $\widehat{L^A}$ and $\widehat{L^C}$. The most traversed countries are Brazil and Argentina. The presence of IXPs in local paths is not so relevant as in other regions, with just 300 paths out of 1\,200 (25\%). 
The IPv6 non-local paths of South America mostly flow through USA, and a very small percentage of them is routed via an IXP. 
The most traversed ASes by non-local paths are transit providers.

\section{Non-locality to infer dependency relationships between countries}
\label{sec:dependency}

\begin{figure*} 
\centering
\begin{tikzpicture}[      
        every node/.style={anchor=south west,inner sep=0pt},
        x=1mm, y=1mm,
      ]   
     \node (fig1) at (0,0)
       {\includegraphics[width=\textwidth]{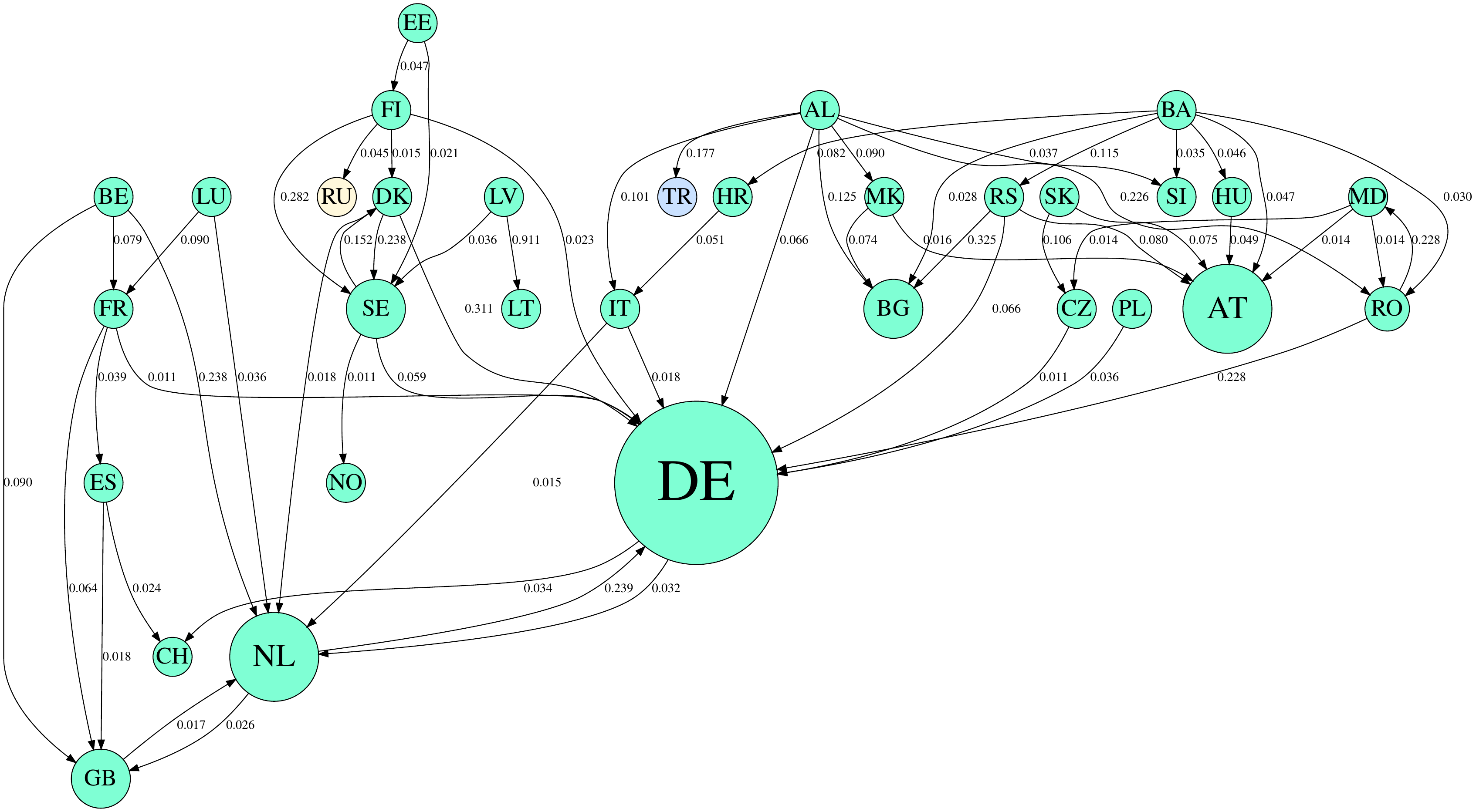}};
     \node (fig2) at (120,3)
       {\includegraphics[scale=0.21,trim=850 460 440 230,clip]{fig/legend_graph.pdf}};  
\end{tikzpicture}
	\caption{Dependency graph across countries in Europe, computed on both IPv6 and IPv4 results; only countries with at least 20 source-destination pairs are considered; only edges with $\widehat{NL^A} \geq$ 0.01 are included; the size of a node $v$ is proportional to $deg^{-}_v$.}
	\label{fig:graph2}
\end{figure*}

We can consider a country dependent on another country
if at least some of the paths having both source and destination in the first do not remain local and go through the second. Dependency relationships can be expressed as a directed graph $(V, E)$, where each vertex $v \in V$ is a country and each edge $e \in E$ is a dependency relationship. An edge $e =(u, v)$ is present if at least some paths that go from $u$ to $u$ are routed via $v$, where $v \neq u$.
Additionally, such edge can be labeled with the related value of $\widehat{NL}$ (this analysis can be performed with both $\widehat{NL^A}$ and $\widehat{NL^C}$ metrics, but in the following we consider only $\widehat{NL^A}$). To compute such value, we compute $\widehat{NL^A}$ for $u$ using $nl(s,d)$. In particular, here $nl(s,d)$ is 1 if a path starting and ending in $u$ traverses $v$, 0 otherwise.
The degree $deg_{v}$ of a vertex $v$ is the number of edges incident to $v$. The in-degree $deg^{-}_{v}$, is the degree calculated by considering only the incoming edges to $v$, and in our metaphor it represents how much such country is a hub of non-local paths for other countries.

To avoid including too weak relationships and/or irrelevant information, 
we add a vertex in $V$ only if for a given country there are at least 20 pairs with source and destination in that country, and we add an edge $(u, v)$ in $E$ only if there are at least two source-destination pairs that show a dependency of $u$ on $v$.
After, we prune edges characterized by values of $\widehat{NL^A} < 0.01$, as this implies that the dependency relationship is limited to a very small portion of the address space. Finally, we removed vertices with $deg = 0$, as the absence of incoming and outgoing edges means that the country does not play any role in the set of dependency relationships.

A dependency graph depicting relationships among countries of different regions is shown in Figure~\ref{fig:graph1}. The color of the vertices encodes the region they belong to, while the size encodes the value of their $deg^{-}$.
Relationships between countries in the same region are not considered: for each edge $(u, v)$, the region of $v$ must be different from the region of $u$.
The sources and the destinations of paths are always in the same country.
The countries with the largest value of $deg^{-}$ are Germany, United States, and United Kingdom (ISO 3166-2: GB). We can also observe two clusters of dependencies, the one of the Middle East on European countries, and the one of South America on the United States.

Note that a direct edge $(u, v)$ between two countries does not necessarily imply that a path originating from $u$ goes directly to $v$ before coming back to $u$. It is possible that other countries between $u$ and $v$ are crossed but that their dependency ($\widehat{NL^A}$) is lower than the adopted threshold. For instance, this situation might arise when there is a set of country-level paths such as ($u$, $c_1$, $v$, $c_1$, $u$) and ($u$, $c_2$, $v$, $c_2$, $u$): in this case, the dependencies $(u, c_1)$ and $(u, c_2)$ could have values of $\widehat{NL^A} < 0.01$, but the aggregate $\widehat{NL^A}$ for $(u, v)$ is greater than 0.01 and the edge is included in the graph.

Figure~\ref{fig:graph2} depicts the detailed situation in Europe, the region that is better covered in terms of measurements.
In this figure, an edge $(u, v)$ is included in the graph when sources and destinations are located in $u$, where $u$ is a country in Europe, and the route passes through $v$, independently from the region $v$ belongs to.
While at regional level Europe shows really high levels of path-locality, where only Turkey and Russia appear in the graph as out-of-region countries (however with low $\widehat{NL^A}$ values), inside the region it is possible to observe several dependencies across countries.
The role of Germany as a hub is confirmed also in this view. At the European level, also the Netherlands and Austria seem to play a relevant role. No single European country is characterized by a dependency value from the USA that is higher than the considered 0.01 threshold. Thus, while there is a small but not negligible amount of paths that are routed via the USA when sources and destinations are located in Europe (but not necessarily in the same country), the same does not apply when considering locality at the country level. Finally, it has to be noticed that the dependency values of European countries on other European countries are generally low, with few exceptions. This means that most European countries show a high degree of intra-country locality.

\section{Discussion}
\label{sec:discussion}

\begin{figure*}[!t]
    \centering
    \subfloat[][Africa.]{
        \label{subfig:scatter-AF}
        \includegraphics[width=0.32\textwidth]{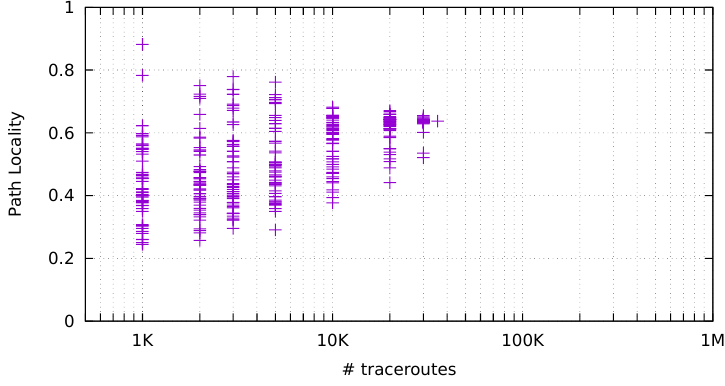}
    }
    \subfloat[][Asia.]{
        \label{subfig:scatter-AS}
        \includegraphics[width=0.32\textwidth]{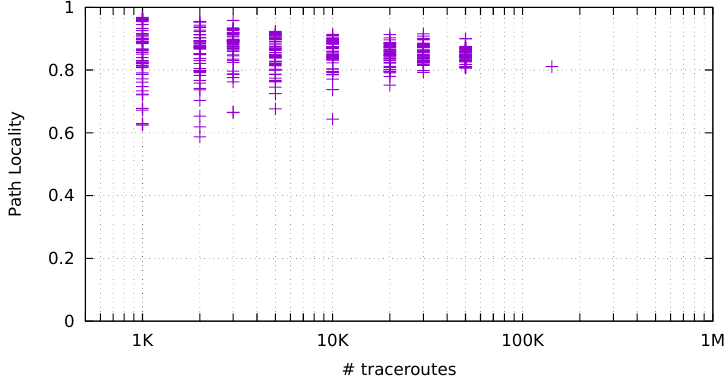}
    }
    \subfloat[][Europe.]{
        \label{subfig:scatter-EU}
        \includegraphics[width=0.32\textwidth]{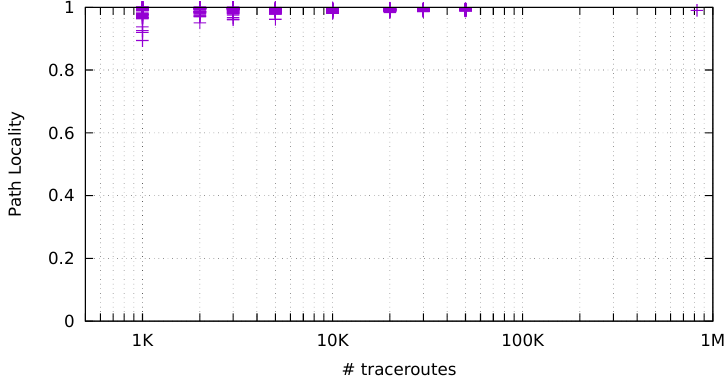}
    }\\
    \subfloat[][Middle East.]{
        \label{subfig:scatter-ME}
        \includegraphics[width=0.32\textwidth]{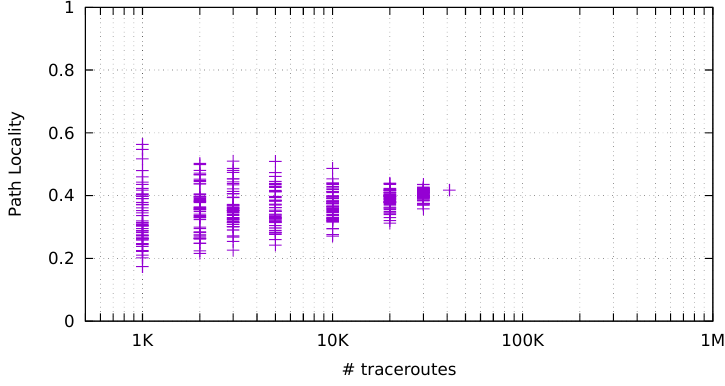}
    }
    \subfloat[][North America.]{
        \label{subfig:scatter-NA}
        \includegraphics[width=0.32\textwidth]{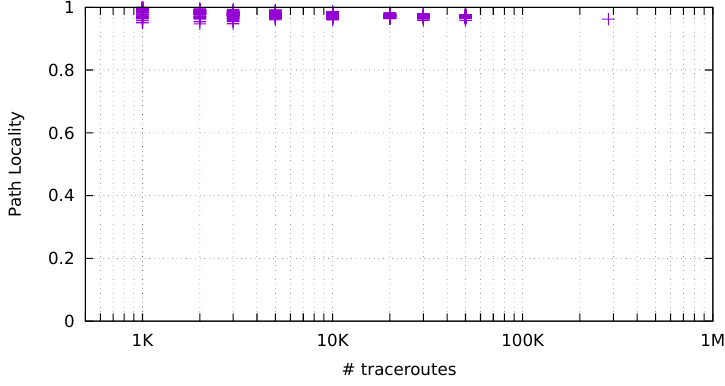}
    }
    \subfloat[][Oceania.]{
        \label{subfig:scatter-OC}
        \includegraphics[width=0.32\textwidth]{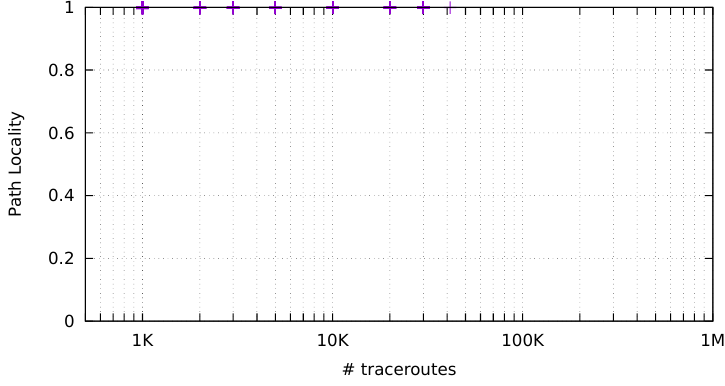}
    }\\
    \subfloat[][South America.]{
        \label{subfig:scatter-SA}
        \includegraphics[width=0.32\textwidth]{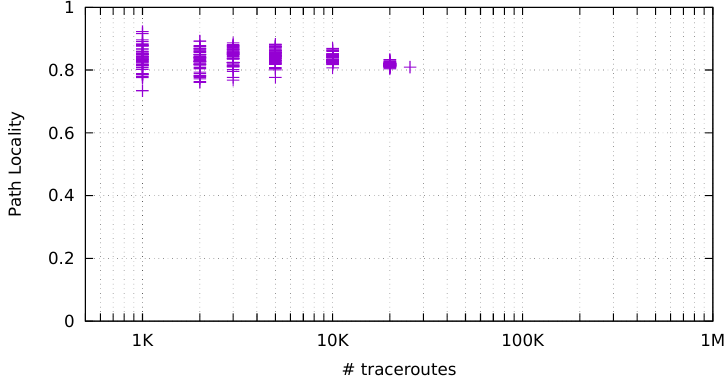}
    }
    \caption{Evolution of $\widehat{L^A_G}$ with increasing numbers of traceroutes.}
    \label{fig:scatters}
\end{figure*}

After having analyzed the situation from multiple points of view, here we provide some general considerations and identify the limitations of the current study. 

\subsection{Considerations on path locality}

The results we presented in the previous sections show that the world is fragmented in terms of locality of Internet paths. Some regions, like Europe, North America, and Oceania, show an almost complete path locality, while some others like Africa and the Middle East show higher levels of non-locality. Asia and South America are somewhere in between. 
In some regions IXPs and Tier-1 providers are able to keep local the great majority of the paths (Table~\ref{tab:topology}), whereas in the other regions external facilities are necessary or preferred. This can be due to a wide range of reasons: partial lack of infrastructure, commercial agreements, limited cross-national coordination. From a geographic point of view, as shown in Table~\ref{tab:reg-countries} and Figure~\ref{fig:graph1}, the attractive force of Europe and North America towards the other regions is evident, at both region and country level, maybe because these are the regions where the Internet initially spread. In addition, in regions characterized by relevant non-locality, the results highlight that $\widehat{L^C}$ is generally less or equal to $\widehat{L^A}$, which means that paths connecting end users are possibly less optimized. This is particularly evident in Africa and South America, where $\widehat{L^C}$ is 0.327 and 0.601, respectively, while $\widehat{L^A}$ is 0.638 and 0.809, respectively.

Table~\ref{tab:topology} shows also that there is not a single recipe for keeping paths local. Some regions mainly rely on local IXPs, like Africa and Oceania, other regions, like North America and South America, rely more on Tier-1 providers. Asia and Europe rely on IXPs, Tier-1 providers, and other providers in almost equal parts, while the Middle East uses mainly local providers with little aid from IXPs and Tier-1 providers. 
In this scenario, it is extremely difficult to give suggestions on how to improve the locality of Internet paths, as the ways to achieve this goal seem numerous and all equally effective. In addition, the technological issues may be just one of the multiple factors that come into play, and other motivations could be equally important.
However, these factors are out of scope for this study, which instead aims at characterizing the recent Internet path locality and providing to the research community a methodology to better evaluate this phenomenon.

\subsection{Limitations}

The results we presented are based on $\sim 1.5$ million traceroutes (derived from an initial dataset of 300+ million), collected by means of a large number of vantage points.
All the vantage points belong to RIPE Atlas, thus the obtained results are shifted towards the view of the Internet that can be obtained from such platform. However, it is important to note that the set of targets is not limited to Atlas nodes, as it includes also hosts that are not part of the measurement infrastructure. In particular, the targets in User Defined Measurements can be generic hosts that the users of the platform considered relevant for their monitoring or data collection purposes.

To better characterize our dataset, in the perspective of analyzing the path locality at a worldwide level, we performed the following analysis. For each region, we selected random subsets of traceroutes of increasing cardinality, and computed the resulting path locality values using such subsets. For each cardinality, we did 50 repetitions. The cardinalities we chose are 1\,000, 2\,000, 3\,000, 5\,000, 10\,000, 20\,000, 30\,000, and 50\,000 traceroutes. For those regions that do not have enough traceroutes, we stopped at the maximum possible value. Results are depicted in Figure~\ref{fig:scatters}. The scatterplots show the $\widehat{L^A}$ values computed for each repetition and each cardinality. In addition, the plots show the $\widehat{L^A}$ value computed with the entire set of traceroutes of each region. The results show that for Europe, North America, and Oceania, the computed $\widehat{L^A}$ values are quite stable also with very small subsets of the original set of traceroutes. This suggests that the values of path locality are not influenced by the specific subset of paths considered. Asia, the Middle East, and South America show larger variability. In Africa, results are even more dispersed, probably because Africa is a vast region, and the number of sources in the region is relatively limited. In addition, as shown in Section~\ref{sec:regionscountries}, Africa is fragmented in terms of locality, and this makes the scatterplot more variable, as the computed path locality value is more dependent on the specific considered subset of paths. Eventually, all the scatterplots necessarily converge to a single point, as that is the path locality value expressed by our dataset. However, the variability in the different scatterplots provides an indication of how much the results are dependent on the collected paths. 

When considering results at the country level, we also adopted a threshold on the minimum number of samples to reduce the possible impact caused by the ones specifically collected. 
In this optic, the relationship among countries and regions depicted in Figure~\ref{fig:graph1} and Figure~\ref{fig:graph2} must be interpreted as non-exhaustive and purely based on the relationships unveiled by our dataset, with the current availability of sources. The method we adopted assumes that a path is local unless it is proven to be non-local. As a consequence, it is possible that we have been unable to capture all the dependencies at the country level just because the limited number of source-target pairs, in some countries, did not allow us to see some existing non-local paths. 

Rather obviously, analyzing the results produced by others measurement platforms would be interesting, in particular to better cover the areas that are under-represented in our study. China is a notable example. Unfortunately, the problem of estimating the locality of paths requires the presence of vantage points in the region of interest. In fact, it is possible to explore multiple paths from a single source, by probing multiple targets in a region, but the source {\em must} be in the same region. It is not possible to estimate the path locality of a region from the outside. Thus, for the specific case of the large Asian country, a better view cannot be obtained by using the measurements originated by other platforms.
Despite the lack of details on some specific parts of the Internet, we believe the overall picture we provided to be valuable in understanding the global situation.

Another limitation of our work is introduced by the incomplete geolocation of the hops of the traceroutes. As described in Section~\ref{sec:dataset}, despite using a considerably accurate geolocation method and having on average almost 70\% of the hops of the traceroutes geolocated, the hops lacking geolocation may impact the accuracy of our inference of locality of the paths.

Finally, the last limitation is that---despite the considerations reported above---to really investigate the causes of non-local paths, an analysis of single source-target pairs would be needed, possibly involving the many Internet operators involved in the process.

\section{Conclusion}

We provided definitions of locality metrics that go beyond the pure fraction of paths that cross the borders of the considered region or country. In particular, we incorporated into the definition a weight that takes into account either the address spaces of sources and destinations or the amount of served population. 
Results show that world regions and countries are characterized by significant differences in terms of path locality. The presence of a large fraction of non-local paths has an impact on the observed end-to-end communication latency, as such routes are particularly circuitous. From a low-level perspective, this information can be useful when planning the deployment of new network infrastructure. 
At a higher level, the most significant dependencies between countries caused by non-local paths have been identified. Some countries---United States, Germany, and United Kingdom---are particularly significant from this point of view. 

Note that the constraints introduced in Section~\ref{sec:dataset} tend to produce conservative results, and that the real amount of path non-locality can be slightly higher than the one we presented. 

\noindent\textbf{Reproducibility.} The measurements used in this study are publicly available at~\cite{atlas}. The measurements we created for this work are easily selectable with the tag ``mcwlm''. All the datasets used for the enrichment are open, details are provided in Section~\ref{sec:dataset}.

\section*{Acknowledgment}

This work is partially funded by the Italian
Ministry of Education and Research (MIUR) in the framework of the CrossLab project (Departments of Excellence). The views expressed are solely those of the authors.

\balance
\bibliographystyle{elsarticle-num}
\bibliography{bibliography}

\end{document}